\documentstyle[prl,aps,epsf,epsfig,floats,twocolumn,amssymb]{revtex}

\newcommand{\be}{\begin{equation}}
\newcommand{\ee}{\end{equation}}
\newcommand{\ba}{\begin{eqnarray}}
\newcommand{\ea}{\end{eqnarray}}
\newcommand{\nn}{\nonumber}

\begin{document}
\draft
\twocolumn[\hsize\textwidth\columnwidth\hsize\csname
@twocolumnfalse\endcsname

\title{Identification of a Scalar Glueball}

\author{M. Albaladejo and J. A. Oller \\}

\address{
 Departamento de F\'{\i}sica, Universidad de Murcia, E-30071 Murcia,
Spain}
\maketitle


\begin{abstract}
We have performed a coupled channel study of the meson-meson S-waves involving isospins
($I$) 0, 1/2 and 3/2 up to 2 GeV. For the first time the 
channels  $\pi\pi$, $K\bar{K}$, $\eta\eta$, $\sigma\sigma$, 
$\eta\eta'$, $\eta'\eta'$, $\rho\rho$, $\omega\omega$, $\omega\phi$, $\phi\phi$,
$a_1\pi$ and $\pi^*\pi$ are considered. 
 All the resonances with  masses below 2 GeV for $I=0$ and $1/2$ are generated by the approach.
 We identify the  $f_0(1710)$ and a pole at 1.6~GeV, which is 
 an important contribution to the $f_0(1500)$, 
 as  glueballs. This is based 
on an accurate agreement of our results with predictions of lattice QCD 
and the  chiral suppression of the 
coupling of a scalar  glueball to $\bar{q}q$.
Another nearby pole, mainly corresponding to the $f_0(1370)$, 
is a  pure octet state not mixed with the glueball.
\end{abstract}

\vspace{1cm}

{\bf 1. } 
QCD, the present theory of strong interactions, is a non-abelian Yang Mills theory 
so that gluons carry colour charge and interact between them. It is generally believed 
that QCD predicts the existence of mesons without valence quarks, the so called 
glueballs. Its confirmation in the spectrum of strong interactions is then at the heart 
of the theory. In quenched lattice QCD 
 the lightest glueball has the quantum numbers of the vacuum, 
$J^{PC}=0^{++}$, with a mass of $(1.66\pm 0.05)$~GeV \cite{specquench}.
 Experimentally the closest  $0^{++}$ scalar resonances to this
energy range are the $f_0(1500)$ and $f_0(1710)$ \cite{pdg}. 
Some references favour 
 the $f_0(1500)$ as the lightest scalar 
glueball \cite{amsler},
 while others do so for the $f_0(1710)$ \cite{weingarten,chanowitz}.  
 
We analyze the $I=0$ meson-meson  S-wave
in terms of 13 coupled channels, $\pi\pi$(1), $K\bar{K}$(2), $\eta\eta$(3),
 $\sigma\sigma$(4), $\eta\eta'$(5), $\eta'\eta'$(6), $\rho\rho$(7), $\omega\omega$(8),
 $K^*{\bar{K}}^*$(9), $\omega\phi$(10) $\phi\phi$(11), 
$a_1(1260)\pi$(12) and $\pi^*(1300)\pi$(13). The number labelling each state is given
  between brackets to the right. The multipion states,
 which play an increasing role for energies above $\sim 1.2$~GeV, are
 mimicked through the $\sigma\sigma$, $\rho\rho$ and $\omega\omega$ channels. 
  It is worth stressing that our
   approach is the first one with such a large number of 
channels and that a similar scheme could also be applied to other controversial 
 meson-meson partial waves. In addition, we study simultaneously the
S-wave of   $K^-\pi^+\to K^-\pi^+$ (involving $I=1/2$ and $3/2$) 
 with the  coupled channels $K\pi$, $K\eta$ and $K\eta'$.


{\bf 2. } 
Let $T_{i,j}^{(I)}$ be the $i\leftrightarrow j$ S-wave amplitude 
with isospin $I$ and $i,j=1\ldots n$, with $n$ 
the number of channels. We use the master formula $T^{(I)}=\left[I+N\cdot g \right]^{-1}\cdot N~,$ 
where $N$ is the symmetric matrix of interaction kernels  and $g$ is a diagonal
matrix of elements $g_i(s)$. 
The function $g_i(s)$ is calculated from kinematics in terms of a
once subtracted dispersion relation and a
 subtraction constant $a_i$ \cite{nd}. Since SU(3) breaking is milder 
 in the vector sector we take $a_7=a_8=a_9=a_{10}=a_{11}$.
The rest of subtraction constants are fitted to data. 
The  matrix elements $N_{i,j}$ consist of the sum of two tree level contributions. The first is a contact 
interaction calculated  from the lowest order Chiral Perturbation Theory Lagrangian,
${\cal L}_2$. The second is due to the exchange of bare resonances in the $s-$channel with the 
 couplings calculated from the lowest order chiral Lagrangian including  
 an octet and
singlet of $0^{++}$ resonances, ${\cal L_S}$ \cite{chpt}. Explicit expressions 
 of $N_{i,j}$ can be found in ref.\cite{nd} for the simplified case of three
channels without including the $\eta_1$ field. We extend these
Lagrangians from SU(3) to U(3) as the $\eta_1$ field is needed to deal with the $\eta$ and $\eta'$
mesons,
similarly as in ref.\cite{jop00}. 
 The  matrix $\Phi=\sum_{i=1}^8 \phi_i
 \lambda_i/\sqrt{2}+\eta_1/\sqrt{3}$
  incorporates in a standard way the nonet of the lightest pseudoscalars.
We also employ the matrix  $U=\exp(i \sqrt{2} \Phi/f)$ 
and the covariant
derivative $D_\mu U=\partial_\mu U -i r_\mu U+i  U \ell_\mu$, with
 $f$ the pion decay constant in the chiral limit fixed to $f_\pi=92.4$~MeV. 
 The classical left and right external  fields, 
  $r_\mu$ and $\ell_\mu$, respectively, are necessary to gauge the global chiral symmetry to a 
  local one \cite{chpt}.
The field $v_\mu=(r_\mu+\ell_\mu)/2$  plays a special role in our approach since it is identified with
$\lambda\, W_\mu$, where $W_\mu$ is the nonet of the lightest $1^{--}$ vector resonances and 
$\lambda$ is a constant, with  $\lambda=4.3$ from 
the width $\rho\to \pi\pi$. The couplings of 
the vector-vector states to the pseudoscalar-pseudoscalar 
 and $\sigma\sigma$ ones are then determined by minimal coupling \cite{ulfreview}. 
Our fits require a singlet and two octets of bare resonances. The two octets were  already 
 considered in ref.\cite{jop00} in the study of  $K^-\pi^+\to K^-\pi^+$.
 We fix the parameters
 of the first octet, mass and coupling constants, to those  in ref.\cite{jop00}, $M^{(1)}_8=1.29$~GeV, $c^{(1)}_d=c_m^{(1)}=26$~MeV.  
 The bare mass of the second octet is fixed  from the same reference, $M_8^{(2)}=1.90$~GeV.  
 We are then left with three parameters for the singlet, $M_1$,
 $\widetilde{c}_d^{(1)}$, $\widetilde{c}_m^{(1)}$, and two for the second octet, 
$c_d^{(2)}$ and $c_m^{(2)}$. It results from our fits that $M_1\lesssim 0.9$ GeV.

Concerning the $\sigma\sigma$ channel we follow a novel method to calculate its transition 
amplitudes, $N_{i,4}$, without including any new free parameter. This can be done 
because the $\sigma$ corresponds to a pole due to the interactions between two pions
in the $I=0$  S-wave, $(\pi\pi)_0$ \cite{ddecays}.  
For the interaction kernel $N_{i,4}$ one starts by  
calculating from the Lagrangians ${\cal L}_2$ and ${\cal L_S}$ 
the tree level amplitude $T^{2+S}_{i,4}$  for  $i\to (\pi\pi)_0(\pi\pi)_0$. 
 To take into account the pion final state interactions, 
 $T^{2+S}_{i,4}$ is multiplied by the factor $\prod_{k=1}^m 1/D(s_k)$, 
with $m$ the number of $\sigma$'s in the scattering process (2 or
 4) and $s_k$ the total centre of mass (CM) energy squared of the $k_{th}$ pair. 
 We use here that the rescattering of two $I=0$ S-wave pions from
 a production kernel is  given by the factor
$1/D=1/(1+V_2 g_1)$, with $V_2=(s-m_\pi^2/2)/f^2$  \cite{ddecays}. To
isolate  $N_{i,4}$  one takes the limit (for definiteness $i\neq 4$)
\be
\lim_{s_1,s_2 \rightarrow s_\sigma} 
\frac{T_{i,4}^{2+S}}{D_{II}(s_1)D_{II}(s_2)}=  \frac{N_{i,4} \,g^2_{\sigma\pi\pi}}{(s_1-s_\sigma)(s_2 -
s_\sigma)}~.
\label{lim}
\ee
 Where 
the subscript $II$ indicates that the corresponding function is calculated on the
second Riemann sheet (with the sign reversed in the definition of the 
pion three-momentum), $s_\sigma$ is the 
$\sigma$ pole position and $g_{\sigma\pi\pi}$ is its coupling to $\pi\pi$. Performing
 the Laurent expansion around $s_\sigma$ of
$1/D_{II}(s)=\alpha_0/(s-s_\sigma)+\ldots$  the evaluation of $N_{i,4}$ 
from eq.(\ref{lim}) requires the ratio
$(\alpha_0/g_{\sigma\pi\pi})^2$.  Since
 $g_{1,II}(s_\sigma)=-f^2/(s_\sigma-m_\pi^2/2)$ at $s_\sigma$, where $1+V_2 g_{1,II}=0$, and 
taking $T_{II}\simeq V_2/(1+V_2 g_{1,II})$, 
 appropriate for these energies \cite{ddecays}, then 
$(\alpha_0/g_{\sigma\pi\pi})^2=f^2/(1-\frac{dg_{1,II}}{ds}
\large{|}_{s_\sigma}\frac{(s_\sigma-m_\pi^2/2)^2}{f^2})\simeq f^2$.
 In this way, $N_{i,4}=T^{2+S}_{i,4} f^2$, $i\neq 4$,  and $N_{4,4}=T^{2+S}_{4,4}
f^4$. Using  $N_{i,4}$ evaluated with $s_k=s_\sigma$  violates unitarity because
$s_\sigma$ is complex and $N_{i,4}$ must be real. Instead, we 
interpret the width of the $\sigma$ resonance 
 as a Lorentzian mass distribution around its nominal mass value $\sim 450$~MeV with a width 
 $\sim 500$ MeV.  
 In this way the $\sigma$ masses ($\sqrt{s_k}$) used to calculate 
  the functions $N_{i,4}$ and $g_{4}$ are folded 
with the previous mass distribution. 
Similarly, for the $\rho\rho$ state $g_7$  is also convoluted with a 
$\rho$ mass distribution. 

\begin{figure}[!ht]
\centerline{\tiny \input{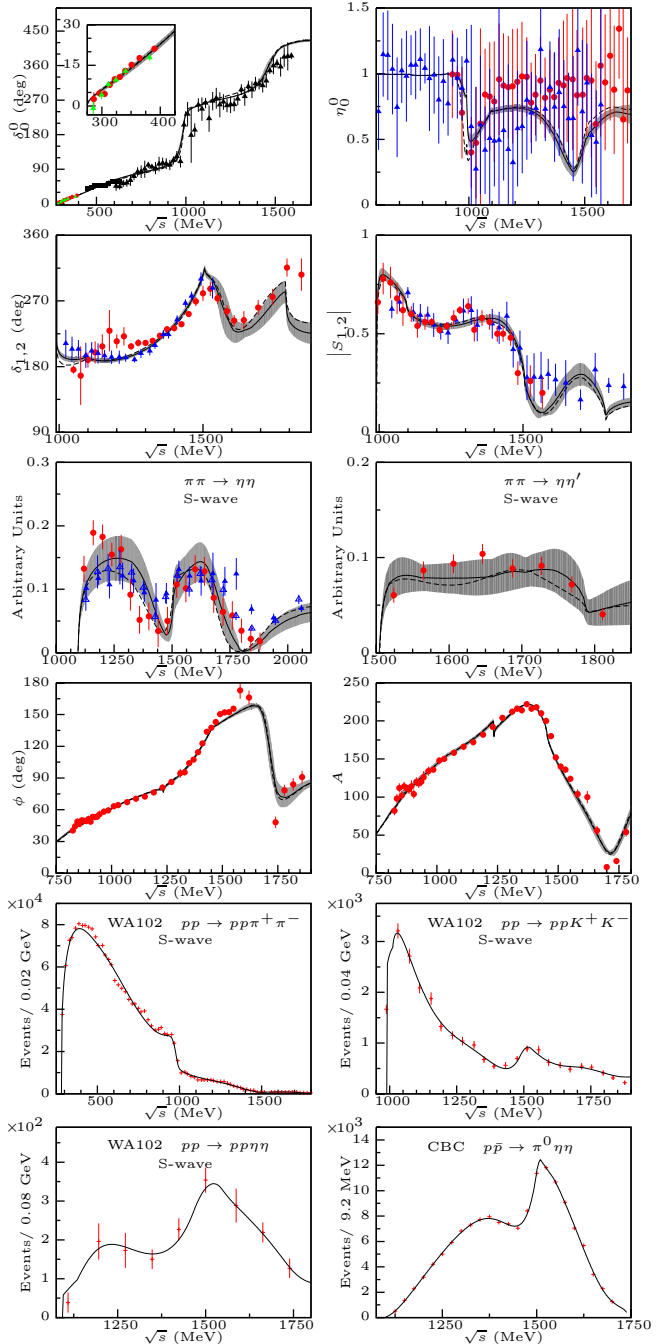}}
\vspace{0.2cm}
\caption[pilf]{{\small Fit to experimental data. More details are given in the text.}  
\label{fig:fit}}
\end{figure}

{\bf 3. }We fit our 12 free parameters to 370 data points from threshold up to  2~GeV. The data 
 comprise the $I=0$ S-wave $\pi\pi$  phase shifts
$\delta_0^0$, the elasticity $\eta_0^0=|S_{1,1}|$, 
the $I=0$ S-wave $\pi\pi\to K\bar{K}$ phase shifts
$\delta_{1,2}$ and modulus  $|S_{1,2}|$, the S-wave contribution to 
the $\pi\pi\to \eta\eta$, $\eta\eta'$ event distributions   
and the phase ($\phi$) and modulus ($A$) of
the $K^-\pi^+\to K^-\pi^+$ amplitude from the LASS data. 
The S-matrix element $S_{i,j}$ is given by
  $S_{i,j}=\delta_{ij}+2i\sqrt{\rho_i}\,T^{(I)}_{i,j}\sqrt{\rho_j}$, where
$\rho_i=q_i/8\pi\sqrt{s}$ and $q_i$ is the CM three-momentum for channel $i$. 
In order, these data are shown 
 on the first eight panels 
 of Fig.\ref{fig:fit} from top to bottom and left to right.
 For  $\sqrt{s}\leq m_K$ in the  $\delta_0^0(s)$ panel 
 we  have
the inset  showing in detail the precise data from $K_{e 4}$ decays. 
The reproduction of the data is fair, as shown in the figure.
 The  dashed lines on the first eight panels 
 include the $a_1\pi$ and $\pi^*\pi$ states, while the solid ones do not.
 The similarity between both curves 
 indicates that these channels give small  
 contributions.
 The width of the band represents our systematic uncertainties at the level of two standard
deviations, $n_\sigma=\Delta \chi^2/(2 \chi^2)^{1/2}$ \cite{etkin}. 
 Compared with other works \cite{ana,polandia,longa}
 we determine the interaction kernels  from standard chiral
Lagrangians, avoid ad-hoc parameterizations,  
include many more channels and fewer free parameters are used. 
For $I=1/2$ the $\kappa$ pole is located at $(708\pm 6-i\,313\pm 10)~$MeV, the $K^*_0(1430)$ at 
$(1435\pm 6-i\,142\pm 8)$~MeV and 
the $K^*_0(1950)$ at $(1750\pm 20-i\,150\pm 20)~$MeV, similarly to ref.\cite{jop00}. 
 For $I=0$ one has the 
 $f_0(600)$ or $\sigma$  at  $(456\pm 6-i\, 241\pm 7)$~MeV and the
 $f_0(980)$ at $(983\pm 4-i\, 25\pm 3)$~MeV.   
 There are poles  at
$(1690\pm 20-i\, 110\pm 20)$~MeV, corresponding to the $f_0(1710)$, and  at  
$(1810\pm 15-i\, 190\pm 20)$~MeV, with mass and width in agreement with those reported 
  for the $f_0(1790)$ by BESII. In the PDG \cite{pdg} the width for the 
$f_0(1710)$  is $137\pm 8~$MeV, much smaller than $220\pm 40$~MeV from the given pole 
position. However, we have checked that on the real axis 
the value of the width corresponding to the half-maximum for the 
partial waves with prominent $f_0(1710)$ peaks 
is just 160~MeV \cite{alba}. This reduction is due to the opening of several channels 
along the resonance region and the agreement with the PDG is restored. 
 The other poles at  $(1466\pm 15 -i\, 158 \pm 12)~$MeV
 and $(1602\pm 15-i 44\pm 15)~$MeV, connected with the $f_0(1370)$ and $f_0(1500)$,
 are referred in the following as 
 $f_0^L$ and $f_0^R$, respectively. Despite that 
 we have included only three bare resonances 
 in $I=0$ we have generated six. 
The poles are located on the unphysical Riemann
sheets that connect continuously with the physical one  for some interval 
along the real $s-$axis.  
Note that the pole $f_0^R$ 
 does not influence the physical axis beyond the $\eta\eta'$ threshold at 1505~MeV, since above
this energy a different Riemann sheet is the one that matches with the physical $s-$axis.
 This effect  typically gives rise to a pronounced signal at the $\eta\eta'$ 
 threshold and this is the reason for the  $f_0(1500)$ mass, $(1505\pm 6)$~MeV \cite{pdg}. 
If a physical amplitude is 
dominated by the $f^R_0$ pole, then its peak at 1505~MeV  has  an 
effective width larger than the one  from the pole position,
 $88$~MeV.  This is so because  
 given a Breit-Wigner located at the position of the $f_0^R$ pole the energy interval 
below 1.5~GeV at which half the value of the modulus squared  at $1.5$~GeV is 
reached  is $\delta=1.2 \Gamma=105$~MeV, 
the width of the  $f_0(1500)$ \cite{pdg}. The $f_0(1370)$ is mainly given by 
the  $f_0^L$ pole, though its precise shape is sensitive to  $f_0^R$ for
 those channels that couple strongly with the latter. In  Fig.\ref{fig:fit} 
 we also show in the last two rows data from $pp$ inelastic scattering at 450 GeV/c
 and $p\bar{p}$ annihilation by the WA102
  and Crystal Barrel (CBC) Collaborations,  respectively. 
 We have fitted the WA102 data using a
 coherent sum of Breit-Wigner functions and a non-resonant term, 
 similarly as done by 
 the WA102 Collaboration \cite{wa102}:
 \ba
&& i) \sqrt{s}< m_\eta+m_{\eta'}~,A=\left\{ \sigma, f_0(980), f_0^L, f_0^R\right\},\nn\\
&&A(\sqrt{s})_i=NR(\sqrt{s})_i+\sum_{j\in A}\frac{a_j e^{i\theta_j}
 g_{j;i}}{M_j^2-s-i\,M_j\Gamma_j}~,
~,\nn\\
&& ii) \sqrt{s}> m_\eta+m_{\eta'}~,B=\left\{  \sigma, f_0(980), f_0(1710), f_0(1790)\right\},\nn\\
&&A(\sqrt{s})_i=NR(\sqrt{s})_i+r_i+\sum_{j\in B}\frac{a_j e^{i\theta_j}g_{j;i}}{M_j^2-s-i\,M_j\Gamma_j}~,\nn\\
&&NR(\sqrt{s})_i=\alpha(\sqrt{s}-m_k-m_\ell)^\beta e^{-\gamma \sqrt{s}-\delta s}~,
\label{wa102} \ea
 where $a_j$ and $\theta_j$ are the amplitude and the phase of the production vertex of the 
$j_{th}$ resonance, $M_j$, $\Gamma_j$ and $g_{j;i}$ are, respectively, the mass, width and 
 the coupling to channel $i$  of the same resonance. The latter is
 determined 
from the residue of the partial waves at the pole position. In addition, 
 $m_k+m_l$ is the threshold for the channel $i$ and
  $\alpha$, $\beta$, $\gamma$, $\delta$ are real parameters.
   The form of the non-resonant term is taken 
 from the WA102 Collaboration \cite{wa102}. The constant $r_i$ is fixed 
  so as the amplitude $A(\sqrt{s})_i$ is continuous at $w_{\eta\eta'}\equiv 
  m_\eta+m_\eta'$. As explained above, once the $\eta\eta'$ threshold is crossed over  
  one has to consider other Riemann sheets 
   which do not have the $f_0^L$ and $f_0^R$ poles 
  but  the $f_0(1710)$ and $f_0(1790)$ ones. 
  Above $w_{\eta\eta'}$ the $\sigma$ and $f_0(980)$
   give tiny contributions. $\Gamma_j$ in eq.(\ref{wa102}) is the largest between 
its value from the pole position and the one calculated 
by summing the partial decay widths $\Gamma_{j;i}= \theta(\sqrt{s}-m_k-m_\ell) \lambda_i
|g_{j;i}|^2 q_i/(8\pi M_j^2)$, with  
$\lambda_i=1/2$  for identical particles. Eq.(\ref{wa102}) incorporates important new facts 
compared to the analyses 
of the WA102 Collaboration. First, the pole positions 
for the different resonances are those already 
determined from our study of the scattering data on the first 8 panels
 of Fig.\ref{fig:fit}. Let us stress that these observables  only involve
two particles in the final state and their analysis is theoretically cleaner.
 Second, the couplings $g_{j;i}$ are similarly fixed. Third, the 
$a_j$ and $\theta_j$ parameters are the same for all the  WA102 
reactions, that 
are fitted simultaneously.
 For the Crystal Barrel data 
on $p\bar{p}$ annihilation we also use eq.(\ref{wa102})
 but without  $NR(\sqrt{s})$.  
A good reproduction of the  data results.
  In  $p\bar{p}\to \pi^0\eta\eta$ one observes 
 a broad bump for the $f_0(1370)$ and a prominent peak for the $f_0(1500)$,
 that also gives strong signals in the WA102 data. Other peaks 
 are observed for the $\sigma$, $f_0(980)$ and 
 $f_0(1710)$. The latter is  important for the 
 the shoulder in $pp\to pp\eta\eta$ above 1.5~GeV.

\begin{table}[ht]
 \begin{center}
\begin{tabular}{|c|c|c|c|}
    GeV               &	 $f_0(1370)$          &     $f_0^R$          &          $f_0(1710)$       \\ 
\hline
$|g_{\pi^+\pi^-}|$      & $3.59\pm 0.16$ & $1.30\pm 0.22$ & $1.21\pm 0.16$   \\
$|g_{K^0{\bar{K}}^0}|$  & $2.23\pm 0.18$ & $2.06\pm 0.17$& $2.0\pm 0.3$     \\ 
$|g_{\eta\eta}|$        & $1.7\pm 0.3$   & $3.78\pm 0.26$ & $3.3\pm 0.8$     \\
$|g_{\eta\eta'}|$       & $4.0\pm 0.3$   & $4.99\pm 0.24$ & $5.1\pm 0.8$     \\
$|g_{\eta'\eta'}|$      & $3.7\pm 0.4$   & $8.3\pm 0.6$   & $11.7\pm 1.6$    
\end{tabular}
\caption{Couplings  of the 
$f_0(1370)$, $f_0^R$ and
$f_0(1710)$. \label{tab:coup}}
\end{center}
\end{table}

{\bf 4. }In table \ref{tab:coup} we give the couplings of the
 $f_0^L$ (identified as the $f_0(1370)$),
  $f_0^R$ and $f_0(1710)$  poles to  the two pseudoscalar channels.
 We observe that the couplings of the  $f_0^R$ and $f_0(1710)$ are quite similar.
This is so  because the two poles coalesce in the same one when moving continuously 
from the sheet of one of them to the one of the other. They
correspond to the same underlying resonance, but split in two 
due to the interaction in coupled channels. 
From the couplings of the $f_0(1710)$ one can calculate the branching ratios 
 $\Gamma(K\bar{K})/\Gamma_{total}=0.36\pm 0.12 (0.38^{+0.09}_{-0.19})$, 
$\Gamma(\eta\eta)/\Gamma_{total}=0.22 \pm 0.12(0.18^{+0.03}_{-0.13})$,
 and 
 $\Gamma(\pi\pi)/\Gamma(K\bar{K})=0.32\pm 0.14(<0.11)$, where the 
 values of the PDG are given 
 between brackets. The values are compatible within one sigma.
We also obtain that 
the $f_0(1790)$  has a small $K\bar{K}$ coupling,
 and this is a major difference with respect to the
$f_0(1710)$ as stressed by BESII. 
  The couplings of the $f_0^L$-$f_0(1370)$ 
in table \ref{tab:coup} correspond to the pure $I=0$ octet member 
 $(\bar{u}u+\bar{d}d-2\bar{s}s)/\sqrt{6}$ because they are 
very close to the tree level  ones $|g_{\pi^+\pi^-}|=3.9$, $|g_{K^0{\bar{K}}^0}|=2.3$,
   $|g_{\eta\eta}|=1.4$,
$|g_{\eta\eta'}|=3.7$, $|g_{\eta'\eta'}|=3.8$~GeV  calculated 
from the Lagrangian ${\cal L}_S$ \cite{chpt}, with  $c^{(1)}_d$,
$c^{(1)}_m$ and $M_8^{(1)}$ given above. 
  We have also checked that this is the case for the $K^*_0(1430)$ resonance which 
  is the $I=1/2$ member of the same  octet.  It follows then
    that the first octet is a
 pure one without mixing with the nearby
 $f_0^R$ and $f_0(1710)$. The $f_0^L$-$f_0(1370)$ couplings imply a large 
width to $\pi\pi$ with 
$\Gamma(f_0(1370)\to 4\pi)/\Gamma(f_0(1370)\to \pi\pi)=0.30\pm 0.12$, in good 
agreement with the  interval $0.10$-$0.25$ given in the recent ref.\cite{buggy}.
 Let us see that the pattern of sizes of the couplings of the $f_0^R$ and $f_0(1710)$  
corresponds to the chiral suppression of the coupling 
of a scalar glueball, $G_0$, to $\bar{q}q$ \cite{chanowitz}. According to  ref.\cite{chanowitz}
  this coupling 
 is proportional to the quark mass, which then implies 
 a strong suppression in the production
 of $\bar{u}u$ and $\bar{d}d$ relative to $\bar{s}s$  from $G_0$. 
With  a pseudoscalar mixing angle $\sin \beta=-1/3$ one has that
 $\eta=-\eta_s/\sqrt{3}+\eta_u\sqrt{2/3}$ and $\eta'=\eta_s\sqrt{2/3}+\eta_u/\sqrt{3}$ with
 $\eta_s=\bar{s}s$ and $\eta_u=(\bar{u}u+\bar{d}d)/\sqrt{2}$. Denoting by $g_{ss}$ the production 
 of $\eta_s \eta_s$, $g_{sn}$ that of $\eta_s \eta_u$ and $g_{nn}$ for 
$\eta_u \eta_u$, 
\ba
g_{\eta'\eta'}&=&2g_{ss}/3+g_{nn}/3+2\sqrt{2}g_{ns}/3~,\nn\\
g_{\eta\eta'}&=&-\sqrt{2}g_{ss}/3+\sqrt{2}g_{nn}/3+g_{ns}/3~,\nn\\
g_{\eta\eta}&=&g_{ss}/3+2g_{nn}/3-2\sqrt{2}g_{ns}/3~.
\label{mixing}
\ea
If the chiral suppression of ref.\cite{chanowitz} operates then 
$|g_{ss}|\gg |g_{nn}|$. This together with the OZI rule suppress the coupling $g_{ns}$.
Taking e.g. the couplings of $f_0^R$ 
  one obtains $g_{ss}=11.5\pm 0.5$, $g_{ns}=-0.2$ and
$g_{nn}=-1.4$~GeV, and the strong suppression is clear. We now
consider the $K\bar{K}$ coupling. A $K^0$ in terms of valence quarks corresponds to
$\sum_{i=1}^3\bar{s}_i u^i/\sqrt{3}$, summing over the colour indices,
 and analogously for the $\bar{K}^0$. The production of a colour singlet $\bar{s}s$ 
 from the $K^0\bar{K}^0$ requires then the combination 
 $\bar{s}_i s^j=\delta_i^j \bar{s}s/3+(\bar{s}_i s^j-\delta_i^j \bar{s}s/3)$, and 
similarly for $\bar{u}_j u^i$.
 As the
production occurs from the colour singlet $\bar{s}s$ source, 
only the configuration $\bar{s}s\,\bar{u}u$
 contributes, picking up a suppression factor of 
1/3. In addition, the coupling $g_{ss}$ 
 has an extra factor 2 compared to that
  of a $\bar{s}s\,\bar{u}u$, because the former contains  two $\bar{s}s$.
 One then expects that the coupling to $K^0\bar{K^0}$ has the absolute value $g_{ss}/6$. 
For the $f_0^R$ and $f_0(1710)$ it results 
 $|g_{K^0{\bar{K}}^0}|\simeq 2$~GeV, in good agreement 
 with 
 table \ref{tab:coup}.  Another resonance 
 with a known enhanced coupling to $\bar{s}s$ is the $f_0(980)$. However,   
the sizes of its couplings to $\eta\eta$, $\eta\eta'$ and $\eta'\eta'$ 
 follow the opposite order to the $f_0(1790)$ and $f_0^R$ cases and 
 all of them are much smaller than the coupling to $K\bar{K}$.  
 Note that quenched lattice QCD \cite{weingarten} 
 establishes that the couplings 
 of the lightest scalar glueball to pseudoscalar pairs in the SU(3) limit scales
  as the quark mass, in support of  the chiral suppression mechanism 
  of ref.\cite{chanowitz}, that we also observe as discussed above. 
 This mechanism  
 also implies 
  that the glueball should remain unmixed. 
  This accurately fits with our previous result  
   that both the $f_0^R$ and $f_0(1710)$ do not mix with the nearby 
  $f_0^L$.
In addition,  the masses of the $f_0^R$ and $f_0(1710)$
 poles are in excellent agreement with the 
 quenched latticed QCD   prediction for the mass of the lightest glueball, 
 $(1.66\pm 0.05)$~GeV.   

{\bf 5. }In summary, we have presented a coupled channel study of the $I=0$, $1/2$ 
meson-meson S-waves from $\pi\pi$ threshold up to  2~GeV with
 13 coupled channels.
 All the $I=0$ and $1/2$ $0^{++}$ 
resonances with masses below 2 GeV have been generated. 
The  $f_0(1710)$ and a pole at 1.6~GeV, which is an important contribution 
to the $f_0(1500)$,  are identified as glueballs. 
Another pole at $(1.466-i\,0.157)$~GeV, mainly corresponding 
to the $f_0(1370)$,  
 is shown to be a pure octet member. 
\vskip 10pt
We thank C. Piqueras for his collaboration. 
Financial support from the grants MEC  FPA2007-6277 and
      Fundaci\'on  S\'eneca  02975/PI/05 and 05113/FPI/06 is acknowledged. 


\end{document}